\documentstyle[12pt,epsfig,prc,preprint,aps]{revtex}

\tightenlines

\begin{document}

\title{Covariant formulation of pion-nucleon scattering
\footnote{Presented at the 16th European Conference on Few-Body Problems
in Physics, June 1-6, 1998, Autrans (France)}}

\author{A. D. Lahiff and I. R. Afnan}

\address{Department of Physics, The Flinders University of South
Australia, GPO Box 2100, \\ Adelaide 5001, Australia}

\maketitle

\begin{abstract}
A covariant model of elastic pion-nucleon scattering based on the
Bethe-Salpeter equation is presented. The kernel consists of
$s$- and $u$-channel $N$ and $\Delta (1232)$ poles, along with
$\rho$ and $\sigma$ exchange in the $t$-channel. 
A good fit is obtained to the $s$- and $p$-wave phase shifts 
up to the two-pion production threshold.
\end{abstract}

\vspace*{0.8cm}

Pion-nucleon ($\pi N$) scattering is an important example of a strong
interaction, and as such plays a significant role in many other
nuclear reactions involving pions, for example pion photoproduction.
Ideally a theory describing $\pi N$ scattering should be derived from Quantum
Chromodynamics (QCD), but since
QCD is not at present solvable for low energies, it is necessary to use a
chirally invariant
effective Lagrangian, where the degrees of freedom are baryons and 
mesons, rather than quarks and gluons.
Following the success of meson-exchange models in describing 
the nucleon-nucleon interaction,
a number of meson-exchange models for $\pi N$ scattering have
been developed over the last few years~\cite{pin}. These models invariably 
begin with an effective
Lagrangian which describes the couplings between the various mesons
and baryons.
The tree-level diagrams obtained from this Lagrangian
are then unitarized in a 3-dimensional approximation to the 
Bethe-Salpeter (BS)
equation~\cite{bs}. Convergence is guaranteed by the introduction of
phenomenological form factors at each vertex. There are an infinite
number of 3-dimensional reductions to the BS equation, and there is
no overwhelming reason to choose one particular approximation over 
any other.
Here we describe a covariant model of elastic $\pi N$ scattering in which
the BS equation is solved without any reduction to 3-dimensions.

In principle, the exact $\pi N \leftarrow \pi N$ amplitude  for a given Lagrangian
can be obtained from the BS equation, with a potential consisting of
all 1- and 2-particle irreducible diagrams, and dressed propagators
in the $\pi N$ intermediate state. Since it is impossible to construct such a
potential, as it would contain an infinite number of diagrams, it is
common practice to truncate this kernel and include only the tree-level
diagrams. Furthermore, if only two-body unitarity is required to be maintained,
then the dressed nucleon propagator is
replaced by a bare propagator with a pole at the physical mass.

Our kernel consists of $s$- and
$u$-channel $N$ and $\Delta (1232)$ exchanges, along with $\rho$ and
$\sigma$ exchange in the $t$-channel. 
We do not include any higher baryon
resonances because these contributions are not expected to be
significant for $\pi N$ scattering below the two-pion
production threshold.
The couplings to the pion field are always through derivative
couplings, as required by chiral symmetry.

There is an ambiguity as to the choice of propagator for
a particle with spin-3/2. The most commonly used propagator is the
Rarita-Schwinger propagator, which is known to have both spin-3/2 as well
as background spin-1/2 components~\cite{rs}. Other forms have been
introduced by Williams~\cite{will} and Pascalutsa~\cite{pas}, which
each have only a spin-3/2 component.
In the present paper we use the  Rarita-Schwinger propagator.

To guarantee the convergence of all integrals, we need to associate with
each vertex a cut-off function. We take this cut-off function to be the
product of form factors that depend on the 4-momentum squared of each
particle present at the vertex~\cite{pin}. Each form factor is
chosen
to be of the form
\begin{displaymath}
f(q^2) = \left({ \Lambda ^2 - m^2
\over \Lambda ^2 - q^2} \right) ,
\end{displaymath}
where $q^2$ is the 4-momentum squared of the particle and $m$ is the
mass. A different cutoff mass $\Lambda$ is used for
each particle.

The $s$-channel pole terms present in the potential become dressed when 
the BS equation is iterated. Therefore
bare coupling constants and masses are used in the $N$ and $\Delta$
$s$-channel pole diagrams in the potential. The bare nucleon parameters are
determined by requiring that in the $P_{11}$ partial wave, there is
a pole at the physical nucleon mass with a residue related to the
physical $\pi N N$ coupling constant. Since the dressed $\Delta$ 
has a width, it would be necessary to analytically continue 
the BS equation into
the complex $s$-plane in order to carry out the renormalization
for the $\Delta$. Rather than doing this we treat the bare $\Delta$ mass 
and the bare $\pi N \Delta$ coupling constant as free parameters. Since
the $P_{33}$ partial wave is dominated by the $s$-channel $\Delta$ pole
diagram, the bare $\Delta$ parameters are essentially fixed by the
$P_{33}$ phase shifts. While only the $s$-channel diagrams in the potential 
are dressed by the ladder BS equation, 
Pearce and Afnan~\cite{pa} have shown that if 3-body unitarity is satisfied, then
the $u$-channel diagrams also become dressed. We approximate this by using
physical masses and coupling constants in the $u$-channel $N$ and $\Delta$
diagrams.

We solve the BS equation by first expanding the nucleon propagator
in the $\pi N$ intermediate state into positive and negative energy components,
and then sandwiching the resulting equation between Dirac spinors. 
This gives two coupled 4-dimensional integral equations which are 
reduced to 2-dimensional integral equations after a partial wave 
decomposition. A Wick rotation~\cite{wick} is carried out in order to obtain 
equations suitable for numerical solution. 
With our choice of form factors,
there is no interference from the form factors when carrying out the
Wick rotation,  provided the
cutoff masses are large enough. The cutoff masses also have to be
chosen large enough so that unphysical thresholds, which are generated by
form factor singularities and propagator singularities pinching the
integration contour, are far above the two-pion production
threshold. More details are given in \cite{la}.

The renormalized $\pi N N$ coupling constant is fixed at its
physical value of
$g_{\pi N N}^2/4 \pi=13.5$. The nucleon renormalization
procedure fixes the bare $N$ parameters. The remaining
parameters are determined in a fit to the phase shifts and scattering 
lengths from the SM95 
partial wave analysis of Arndt et al~\cite{SM95}. For the cutoff masses 
we obtain $\Lambda_{N}=3.12$, $\Lambda_{\pi}=1.73$, $\Lambda_{\rho}=4.67$, 
$\Lambda_{\Delta}=4.86$, and $\Lambda_{\sigma}=1.4$ (all in GeV).
The coupling
constants are $g_{\rho \pi \pi} g_{\rho NN} / 4 \pi = 3.2$,
$\kappa_{\rho}=2.57$, 
$g_{\sigma \pi \pi}g_{\sigma N N} / 4 \pi = -0.3$,
$f_{\pi N \Delta}^2/4 \pi = 0.44$, and
$f_{\pi N \Delta}^{(0)2}/4 \pi = 0.35$.
The remaining masses are
$m^{(0)}_{\Delta} = 2.13$ GeV and $m_{\sigma} = 700$ MeV. With these parameters,
the renormalization procedure gives
$m_N^{(0)}=1.37$ GeV and $g_{\pi N N}^{(0)2}/4 \pi = 3.8$.

\begin{figure}
\hspace*{-1.8cm} \vspace*{2cm}
\epsfig{figure=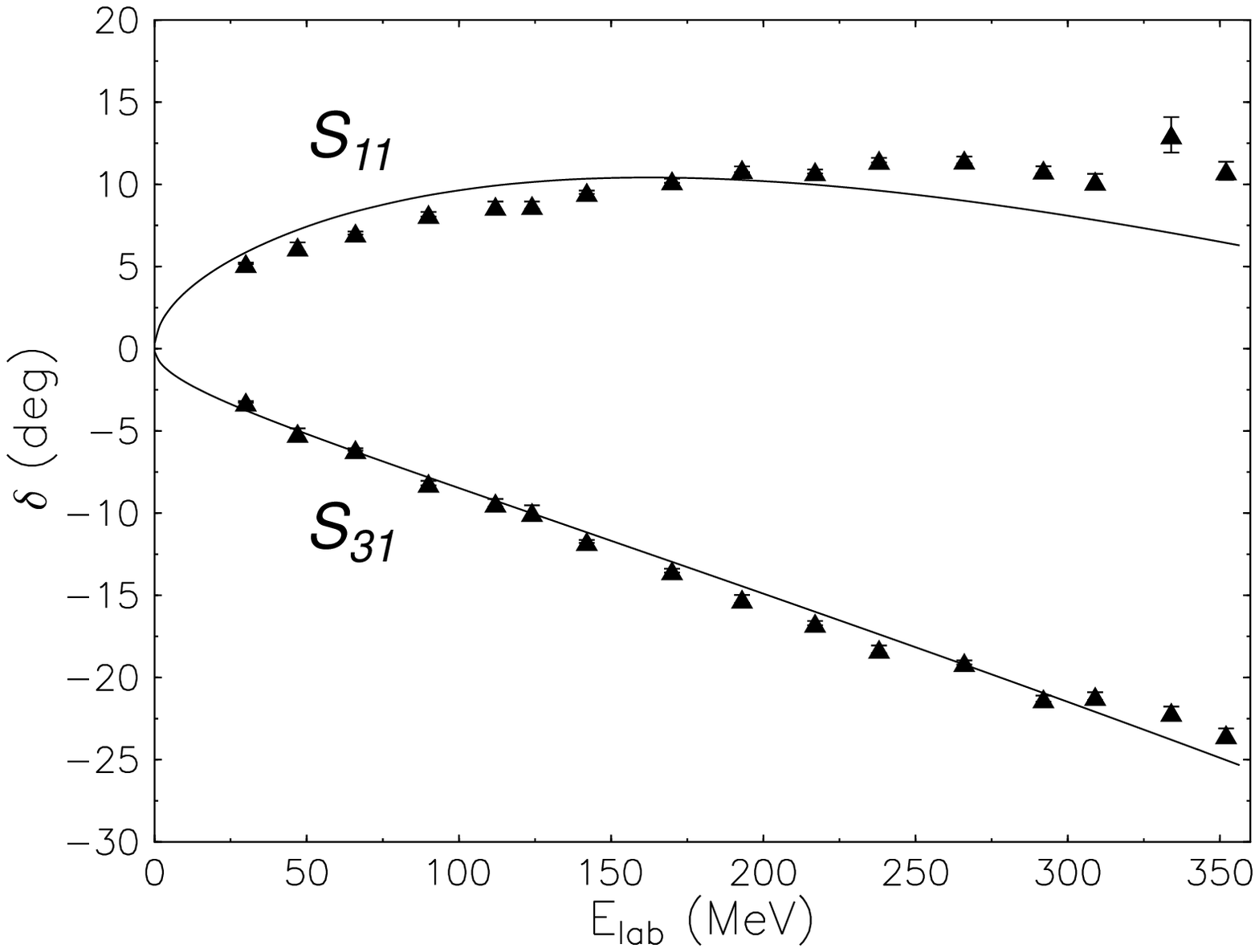,width=9.5cm,scale=1.0}
\hspace*{-1.8cm}
\epsfig{figure=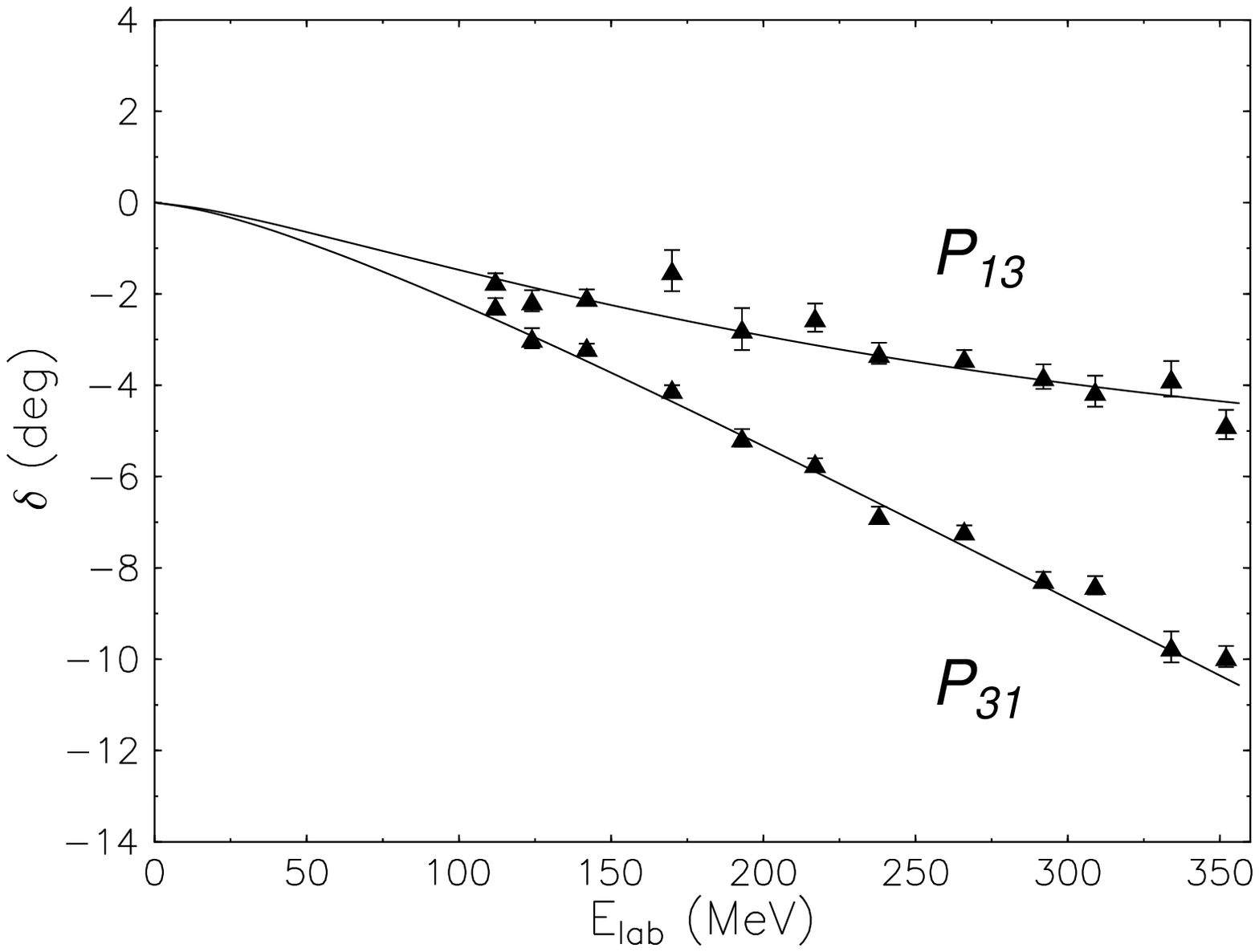,width=9.5cm,scale=1.0}

\vspace*{-2.4cm}

\hspace*{-1.8cm} \vspace*{2cm}
\epsfig{figure=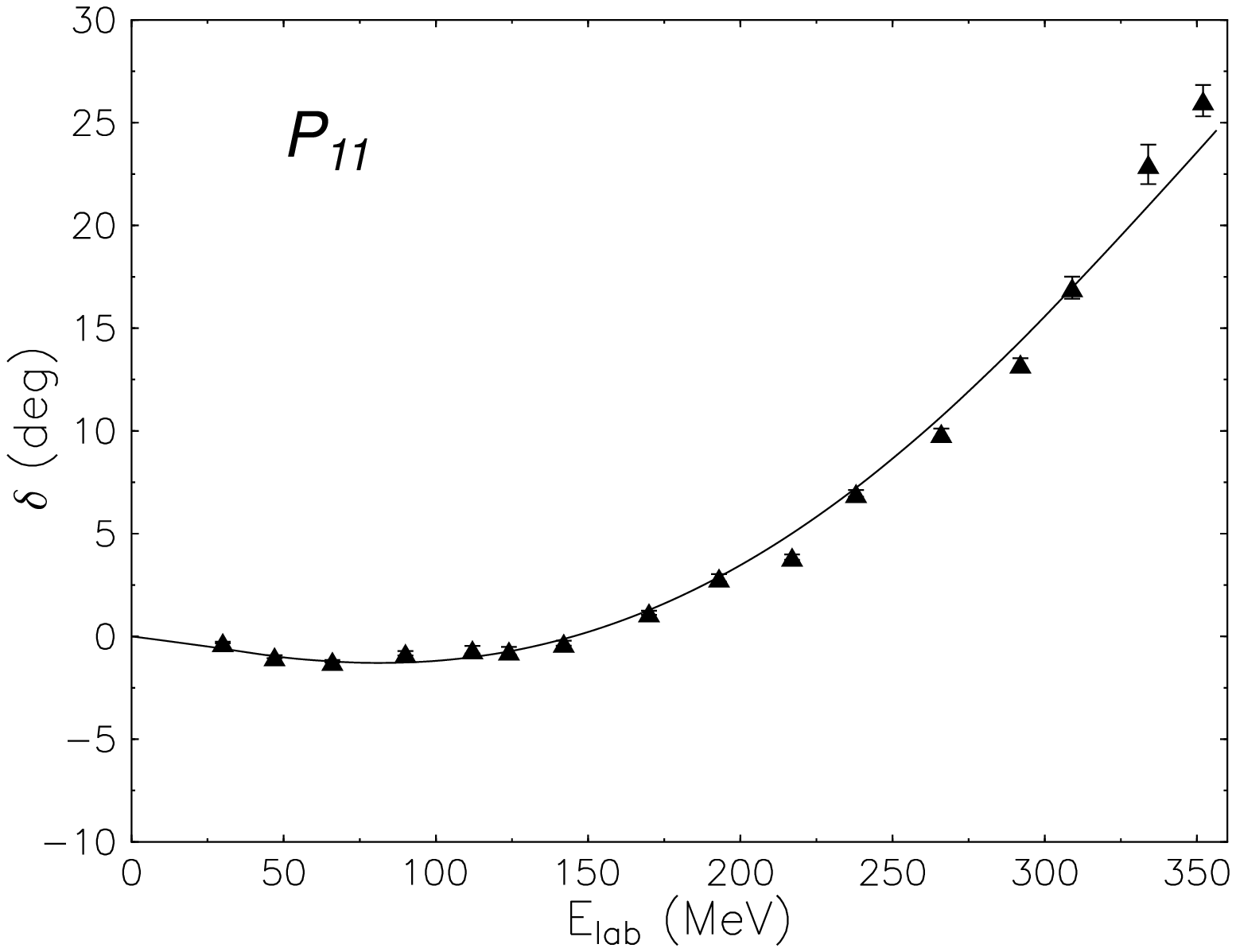,width=9.5cm,scale=1.0}
\hspace*{-1.8cm}
\epsfig{figure=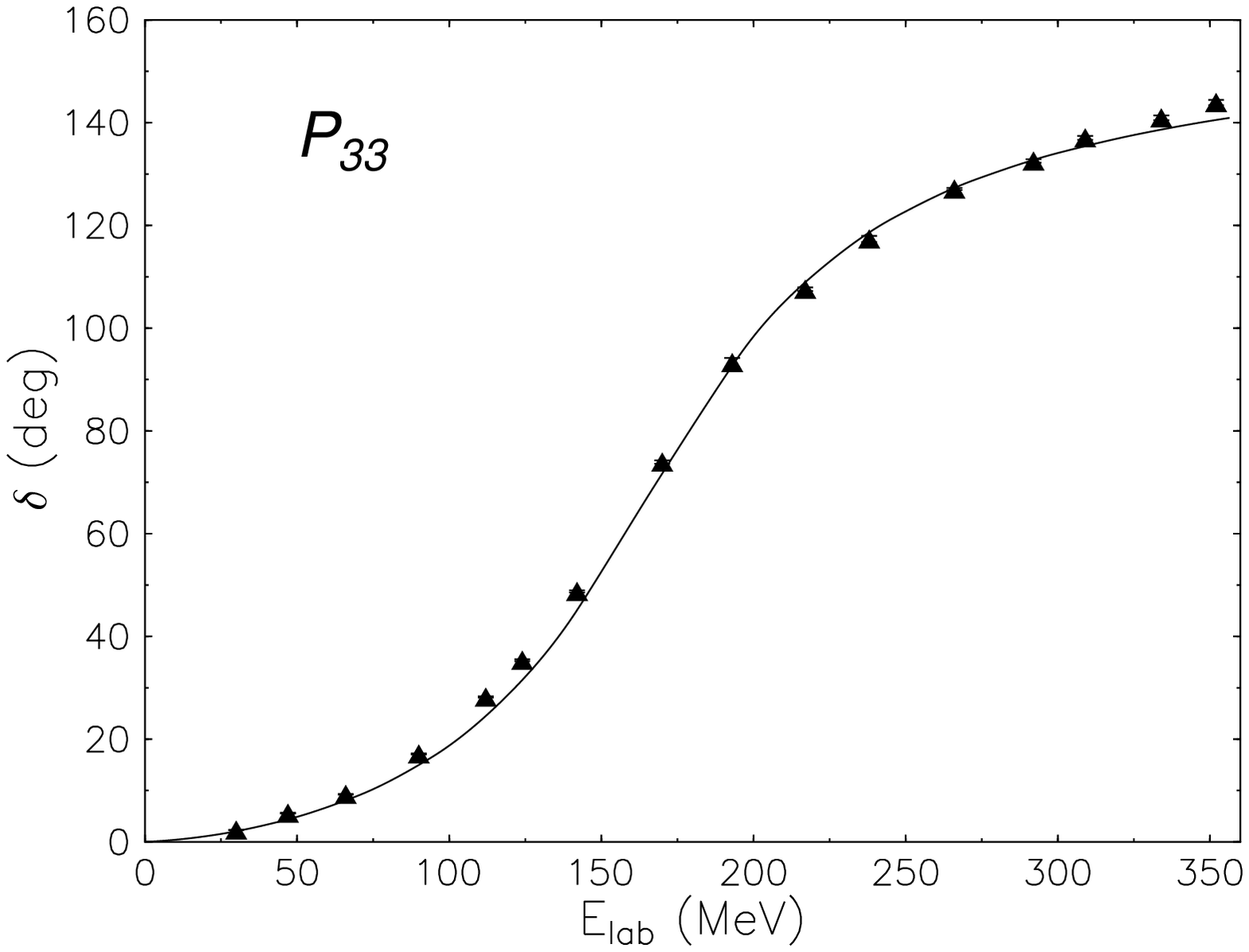,width=9.5cm,scale=1.0}

\caption{The phase shifts obtained from the Bethe-Salpeter equation
are shown versus the pion laboratory energy, compared to the
 VPI SM95 partial wave analysis.}
\label{fig:ps}
\end{figure}

We obtain a good fit to the $s$- and $p$-wave phase shifts. The resulting 
phase-shifts are
shown in Figure~\ref{fig:ps}, and the scattering lengths 
and volumes are shown in
Table~\ref{lengths}. In general our coupling constants are consistent
with those used in the other $\pi N$ models.
Assuming universality ($g_{\rho} \equiv g_{\rho \pi \pi} = g_{\rho NN}$),
our value of $g_{\rho}^2 / 4 \pi$ is close to
that obtained from the decay $\rho \rightarrow 2 \pi$, i.e.
$g_{\rho}^2 / 4 \pi=2.8$, and
$\kappa _{\rho}$ is close to the value $\kappa _{\rho}=3.7$
arising
from vector meson dominance. 
Our physical $\pi N \Delta$ coupling constant is
slightly larger than the value $f_{\pi N \Delta}^2/ 4 \pi = 0.36$
obtained from calculations of the width of the 
$\Delta$ 
(the use of the larger coupling was necessary in order to obtain a good fit 
to the $P_{13}$ and $P_{31}$ phase shifts).
The contribution of $\sigma$-exchange is very small, while the 
$u$-channel $\Delta$ diagram provides a large amount of attraction
in all partial waves except $S_{31}$ and $P_{33}$. The attraction
in the $P_{11}$ partial wave is dominated by  $\rho$-exchange and the 
$u$-channel $\Delta$. Notice that
some additional attraction is required for high energies in the
$S_{11}$ partial wave.

The cutoff masses turn out to be quite large, with the result that the
dressing is significant, as is evident from the large size of the
bare $N$ and $\Delta$ masses. The baryon self energies are dominated by
the one-pion loop diagrams. 
In view of the significance of the dressing,
it is interesting to examine the effect of
the dressing on the $\pi N N$ form factor.
We can calculate a renormalized cutoff mass
by comparing the dressed $\pi NN$ vertex, with the nucleons on-mass-shell
and the pion off-shell, to a monopole form factor.
We find $\Lambda _{\pi}^R=1.17$ GeV (recall that the bare cutoff mass
is 1.73 GeV).
Therefore the vertex dressing softens the $\pi N N$ form factor.

\newpage

\begin{table}
\begin{center}
\begin{tabular}{lcrcrcc|cclcrcr} 
{ } & { } & BSE & { } & SM95 & { } & { } &
{ } & { } & { } & { } & BSE & { } & SM95 \\ \hline
$S_{11}$ & { } & 0.184 & { } & 0.175   
& { } & { } & { } & { } & $S_{31}$ & { } & -0.101 & { } & -0.087 \\
$P_{11}$ & { } &  -0.079 & { } & -0.068 &
 { } & { } & { } & { } & $P_{31}$ & { } & -0.045 & { } & -0.039 \\
$P_{13}$ & { } & -0.037 & { } & -0.022 &
 { } & { } & { } & { } & $P_{33}$ & { }  & 0.181 & { } & 0.209 

\end{tabular} 
\end{center}

\caption{Scattering lengths and volumes obtained from the
Bethe-Salpeter equation (BSE). Units
are $m_{\pi} ^{-(2 \ell+1)}$.}
\label{lengths}
\end{table}

\newpage

\end{document}